Demonstration of electron filtering to increase the Seebeck coefficient in ErAs:InGaAs/InGaAlAs superlattices

J. M. O. Zide[1], D. Vashaee[2], G. Zeng[3], J. E. Bowers[3], A. Shakouri[2], A. C. Gossard[1]

1. Materials Department, University of California, Santa Barbara, California 93106-5050

2. Jack Baskin School of Engineering, University of California, Santa Cruz, Calfornia 95064

3. Department of Electrical and Computer Engineering, University of California, Santa Barbara, CA 93106

Abstract:

In this letter, we explore electron filtering as a technique to increase Seebeck coefficient and the thermoelectric power factor of heterostructured materials over that of the bulk. We present a theoretical model in which Seebeck coefficient and the power factor can be increased in an InGaAs based composite material. Experimental measurements of the cross-plane Seebeck coefficient are presented and confirm the importance of the electron filtering technique to decouple the electrical conductivity and Seebeck coefficient to increase the thermoelectric power factor.

Traditionally, research on thermoelectric materials has focused on finding or synthesizing materials which have a good balance of the relevant properties rather than systematic studies on how to decouple these parameters using new physical approaches. The efficiency of thermoelectric materials is quantified by the dimensionless figure of merit $ZT=S^2\sigma T/\beta$, where S is the Seebeck coefficient, $\sigma$ is electrical conductivity, $\beta$ is thermal conductivity, and T is temperature. Good thermoelectric materials have high electrical conductivity and Seebeck coefficients and low thermal conductivities, but these properties are not independent. In particular, the electrical conductivity and Seebeck coefficient are inversely related, so it is not generally possible to increase the thermoelectric power factor above a particular optimal value for a bulk material. Traditional thermoelectric materials, such as bismuth telluride ($Bi_2Te_3$) and lead telluride (PbTe) have ZT values of ~1. Pioneering work by Hicks and Dresselhaus [1,2] suggested an improvement in thermoelectric power factor for low dimensional solids. Recent work by Venkatasubraminian *et al.* [3] and Harman *et al.* [4] using superlattices and quantum dots based on these materials has increased room temperature ZT to ~2-2.4. These improvements in performance are primarily a result of a reduction in lattice thermal conductivity, and the thermoelectric power factor ($S^2\sigma$) is largely unchanged. Shakouri *et al.* first proposed the use of solid state thermionics for thermoelectric applications [5], which allows the use of materials with high electrical conductivity and low thermal conductivity while relaxing the need for a high Seebeck coefficient. Mahan *et al.* first proposed multilayer thermionic emission to increase the efficiency (ZT) [6] but later concluded that the main advantage of heterostructures was in the reduction of phonon thermal conductivity [7]. Shakouri *et al.* later suggested that tall barriers and highly degenerately doped superlattices could achieve substantial increases in thermoelectric power factor over bulk materials [8]. Vashaee *et al.* revisited electron transport perpendicular to the barrier and determined that highly degenerately doped semiconductor or metal superlattices could achieve power factors higher than the bulk and determined that nonconservation of transverse momentum can have a large effect (especially in the case of metal superlattices) by increasing the number of electrons contributing to conduction by thermionic emission [9]. In this paper, we calculate the electrical and thermoelectric properties of superlattices consisting of degenerately doped $In_{0.53}Ga_{0.47}As$ with $In_{0.53}Ga_{0.28}Al_{0.19}As$ barriers. Experimental measurements show a corresponding increase in room temperature cross-plane Seebeck coefficient. We believe this is direct experimental evidence of the importance of this electron filtering technique to increase the thermoelectric power factors of heterostructures over those of constituent bulk materials.

If barriers are incorporated into a bulk material to create a superlattice, the electron transport properties can be changed drastically. If these barriers are sufficiently thick, the electron motion through the semiconductor will be limited to those electrons with sufficiently high energy for thermionic emission over the barrier. In this type of structure, it is possible to dramatically increase the Seebeck coefficient with a relatively modest decrease in electrical conductivity.

Mathematically, we can write the transport coefficients based on the Boltzman

transport equation as:

$$\sigma = \int \sigma(E)dE \qquad (1)$$

and

$$S = \frac{1}{qT}\frac{\int \sigma(E)(E-E_F)dE}{\int \sigma(E)dE} \qquad (2)$$

where

$$\sigma(E) = q^2 \tau(E) v^2(E) D(E)\left(\frac{-df}{dE}\right), \qquad (3),$$

τ(E) is the energy dependent relaxation time which can be modified to take into account the quantum mechanical transmission probability T(E), v(E) is electron velocity, D(E) is the density of states, and f(E) is the Fermi-Dirac distribution function. If the barriers of a heterostructures are sufficiently thick to neglect the contribution of tunneling, it is possible to use a Heaviside step function approximation: $T(E) \approx \Theta(E-E_B)$, where $E_B$ is the barrier energy and $E=\hbar^2 k^2/2m$. If transverse momentum (i.e. momentum in the xy-plane where the z-direction is the periodic direction of the superlattice) is conserved (i.e. no elastic scattering of momentum in the xy-plane into the z-direction), then $k^2=k_z^2$ and only electrons with sufficient momentum perpendicular to the barriers contribute to the conduction. If transverse momentum is not conserved, motion in the xy-plane is coupled to motion in the z-direction and $k^2=k_x^2+k_y^2+k_z^2$, resulting in a larger differential conductivity. Vashaee *et al.* have shown theoretically that nonconservation of transverse momentum can greatly increase thermionic emission current and, comparing their model to experimental current-voltage characteristics of various superlattice structures, found that transverse momentum is conserved for electrons in planar semiconductor barriers [9,10]. The conservation of transverse momentum is a consequence of the relative translational invariance in the plane of a conventional quantum well. By controlling the roughness of the superlattice interfaces during the growth or by taking advantage of well designed quantum dot structures, it is possible to break this translational invariance and increase the thermionic current density and consequentially, the electrical conductivity.

In this work, we present a model for InGaAs wells with InGaAlAs barriers with a Ga/Al ratio chosen for a barrier height of $E_B$=200meV. Parameters used for these calculations are given in Table I. Briefly, a conventional bulk transport was modified to calculate the in-plane and cross-plane thermoelectric characteristics of the superlattice structures [10]. The quantum mechanical transmission coefficient for electron transport between wells was introduced in the Boltzmann equation to take into account both tunneling and thermionic emission. The contributions in transport from 2D states in the well and 3D states above the well and in the barrier region are separately considered. Since the barrier layer thickness is large enough for miniband transport to be negligible, transmission probability is calculated using the Wentzel-Kramer-Brillouin (WKB) approximation. Finally, Fermi-Dirac statistics were used to account for the number of available empty states in the neighboring wells for tunneling and thermionic emission calculations.

Calculations are performed assuming both conservation and nonconservation of transverse momentum. Room temperature results are given in Fig. 1. Of the increase in ZT, approximately 65% is due to electron filtering, while the remaining 35% is due to reduced thermal conductivity because of phonon scattering by the ErAs nanoparticles in the InGaAs which has been experimentally measured [11,12]. In Fig. 2, similar results are presented at higher temperatures, where the thermoelectric figure of merit for these materials is expected to be higher. In this case, approximately 84% of the increase in ZT is due to electron filtering, while the remainder is due to decreased thermal conductivity.

To test this model experimentally, several $In_{0.53}Ga_{0.47}As/In_{0.53}Ga_{0.28}Al_{0.19}As$ samples (lattice-matched to InP) were grown on both (100) semi-insultating InP:Fe and (100) n-type InP:Sn substrates using a Varian Gen II molecular beam epitaxy (MBE) system. In each case, the substrate temperature was measured at 490°C using a pyrometer. A schematic representation of the sample structure is given in Fig. 3. In each sample, the 20nm InGaAs regions contain 0.3% atomic of co-deposited ErAs nanoparticles. ErAs is a semimetal which forms self-assembled nanoparticles in a rocksalt structure when incorporated into MBE-grown InGaAs [13]. The incorporation of ErAs into semiconductors can have effects including the creation of a buried Schottky barrier [14], the creation of deep states for rapid carrier recombination [15], enhanced tunneling [16], doping [17], or phonon scattering [18,19] to reduce thermal conductivity [11,12]. In this work, ErAs is included as an electron donor and a phonon scattering center for reduced thermal conductivity. Co-doping with ErAs in addition to silicon makes it possible to achieve a large bulk doping ($10^{19}$ cm$^{-3}$) without special efforts and generally results in a larger mobility than InGaAs which is equivalently doped with silicon. Also, it is hoped that the incorporation of ErAs into the InGaAs will result in nonconservation of transverse momentum. The effects of ErAs on the thermoelectric properties of InGaAs and details of the growth of the InGaAs region are detailed elsewhere [20].

The barrier layers consist of 10nm of $In_{0.53}Ga_{0.28}Al_{0.19}As$, which is grown as a digital alloy, consisting of $(In_{0.53}Ga_{0.47}As)_{0.6}(In_{0.52}Al_{0.48}As)_{0.4}$. A digital alloy is a short period superlattice (~1nm period) in which miniband formation results in an effective barrier height which is a weighted average of the two materials. In this case, the effective barrier height is ~200meV.

The InGaAs regions are co-doped with varying amounts of silicon, resulting in samples with InGaAs electron concentrations of 2, 4, 6, and $10\times10^{18}cm^{-3}$. The electron concentration of these samples was measured using the samples grown on semi-insulating substrates using room temperature Hall measurements in a van der Pauw geometry. The free electron concentrations, obtained by dividing the measured sheet carrier densities by the total thickness of InGaAs in the structure, and the corresponding electron mobilities are plotted in Fig. 4.

Using the samples grown on semi-insulating substrates, the room temperature in-plane Seebeck coefficient was measured for each effective doping concentration. A thermoelectric cooler was used to create a temperature gradient across a small sample of each structure. Metal contacts were deposited and the differences in voltage and temperature (the ratio being the Seebeck coefficient) were measured. The in-plane Seebeck coefficients are plotted in Fig. 5 and compared to theoretical values for bulk ErAs:InGaAs obtained using the model we have described.

The cross-plane Seebeck coefficient, which should be larger than the in-plane Seebeck coefficient because of the electron filtering effect, was measured using the samples grown on n-type substrates, in which the superlattice thickness was 2.1μm. Mesas were etched to isolate the superlattice, and metal contacts were deposited on the mesa. A dielectric layer was deposited on top of the structure and a microheater was fabricated. The same structures were fabricated on the substrate (with the superlattice etched away) and the combined Seebeck coefficient of the superlattice and substrate were measured by measuring the voltage and temperature difference between the two structures. The Seebeck coefficient of the substrate was measured by a similar technique, and the Seebeck coefficient of the superlattice was extracted. The details of this measurement technique and the accompanying analyses have been presented elsewhere [21]. The resulting cross-plane Seebeck coefficients are plotted in Fig. 5 and compared to the theoretical values for electron filtering both with conserved and with nonconserved transverse momentum. The measured Seebeck values agree reasonably well with theoretical values, though it is impossible to draw conclusions as to whether transverse momentum is conserved on the basis of these measurements.

Thus, we have shown that electron filtering results in an increase in Seebeck coefficient by a factor of two to three and demonstrated the importance of this technique to increase both the thermoelectric power factor and ZT of heterostructured materials over that of the bulk. This is accomplished by changing the differential conductivity of the material (as a function of electron energy) to decouple the electrical conductivity and the Seebeck coefficient. This approach is compatible with the reduction of thermal conductivity due to phonon scattering by ErAs nanoparticles. Electron filtering offers both a viable approach to more efficient thermoelectric materials and opportunities to obtain new insights into electron and heat transport in semiconductor heterostructures.

This work was supported by the Office of Naval Research as part of the Thermionic Energy Conversion MURI monitored by Dr. Mihal E. Gross.

Fig. 1. Room temperature modeling of cross-plane electrical and thermoelectric properties: (a) electrical conductivity, (b) Seebeck coefficient, and (c) ZT. Solid curves are bulk InGaAs, while dashed curves are superlattices with conserved transverse momentum and dotted curves are superlattices with nonconserved transverse momentum.

Fig. 2. High temperature (900K) modeling of cross-plane electrical and thermoelectric properties: (a) electrical conductivity, (b) Seebeck coefficient, and (c) ZT. Solid curves are bulk InGaAs, while dashed curves are superlattices with conserved transverse momentum and dotted curves are superlattices with nonconserved transverse momentum.

Fig. 3. Schematic diagram of the sample structure. Each sample structure was grown on both semi-insulating substrates for in-plane measurements (30 repeat units) and on conductive (n-type) substrates (70 repeat units) for cross-plane measurements. Samples were doped with both ErAs and varying amounts of silicon.

Fig. 4. Room temperature, in-plane electrical properties of the ErAs:InGaAs wells as a function of ErAs doping plus silicon co-doping. Electron concentration; solid line, filled circles and electron mobility; dashed line, filled squares.

Fig. 5. Room temperature Seebeck coefficient; comparison of experimental and theoretical data in the in-plane (experimental data: squares, theory: solid curve) and cross-plane directions (experimental data: circles, theory: dashed curve for superlattice

with conserved transverse momentum and dotted curve for superlattice with nonconserved transverse momentum). Experimental results indicate an improvement in Seebeck coefficient due to electron filtering.

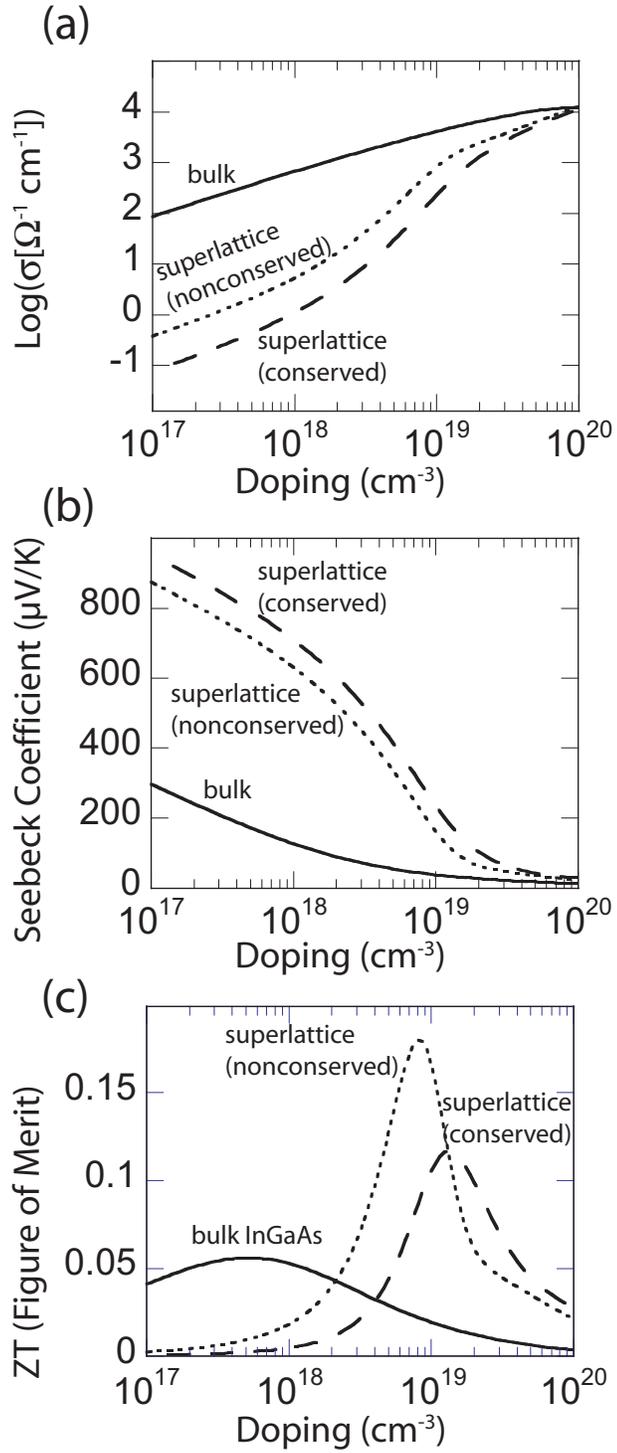

Figure 1, J. M. O. Zide

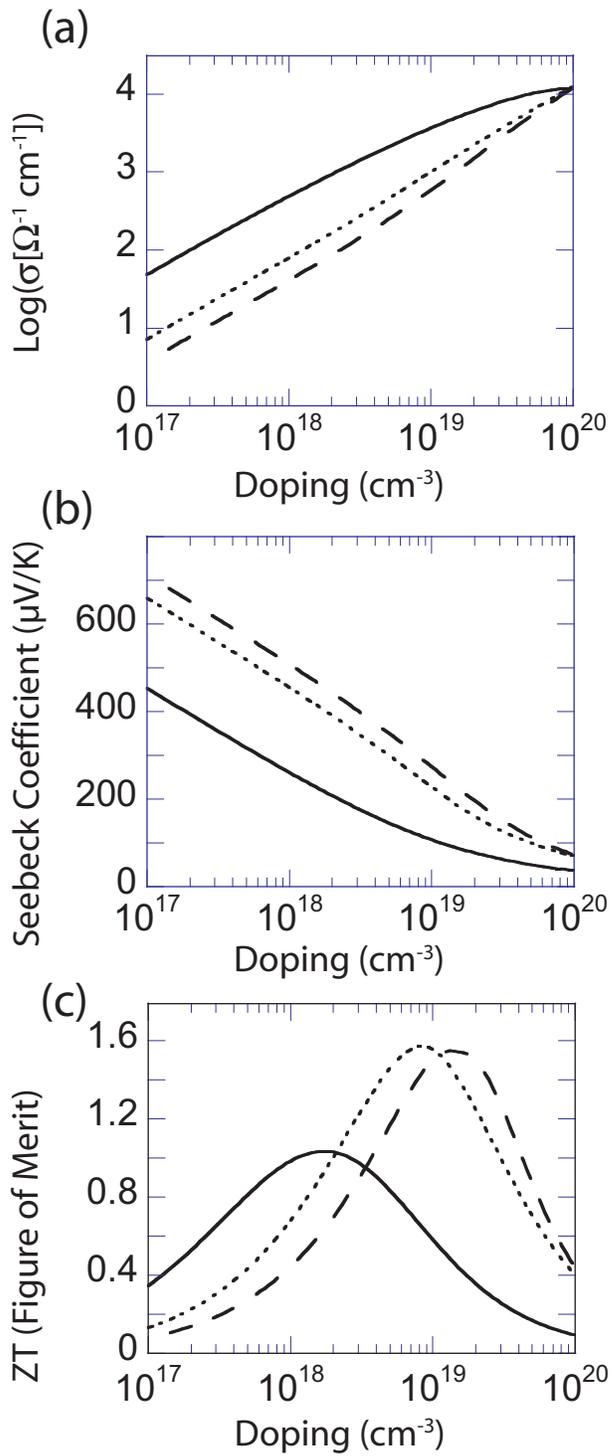

Figure 2, J. M. O. Zide

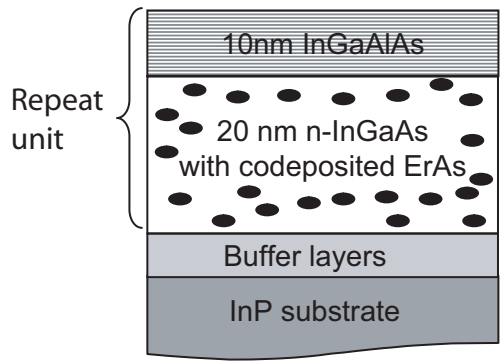

Figure 3, J. M. O. Zide

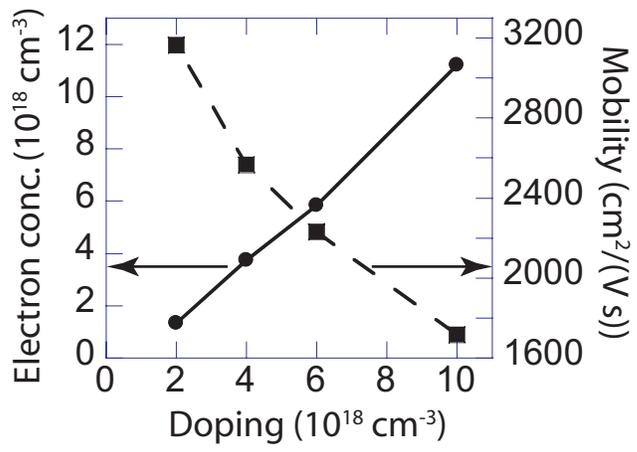

Figure 4, J. M. O. Zide

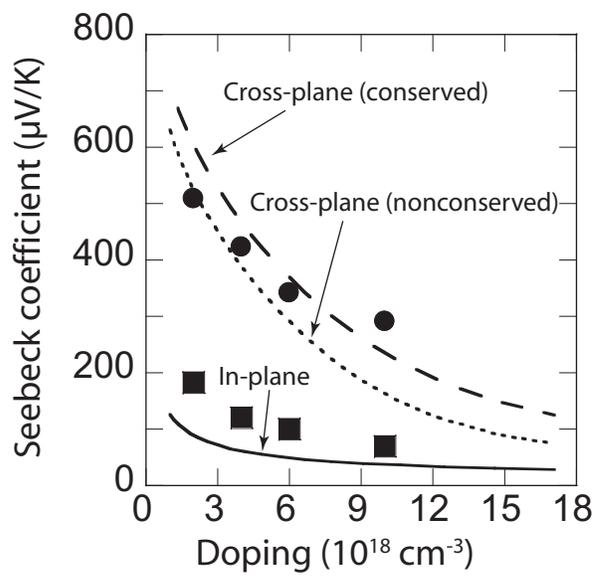

Figure 5, J. M. O. Zide

| Well/Barrier | $n_w$ | $L_w$ (nm) | $L_b$ (nm) | $E_b$ (meV) | $m_{eff}$ (well/barrier) | β (300K) (W/mK) (bulk/SL) | β (900K) (W/mK) (bulk/SL) | $\mu_b$ (cm$^2$/Vs) | $V_s$ (cm/s) | α (eV$^{-1}$) |
|---|---|---|---|---|---|---|---|---|---|---|
| In$_{0.53}$Ga$_{0.47}$As/ In$_{0.53}$Ga$_{0.28}$Al$_{0.19}$As | 70 | 20 | 10 | 200 | 0.043/0.058 | 5.5 / 3.5 | 2.5 / 2.1 | 5600 | 2×10$^7$ | 1.167 |

**Table I**: Parameters used in the simulations of InGaAs/InGaAlAs superlattice, where $n_w$ is the number of wells. $L_w$ and $L_b$ are well and barrier widths, respectively. $E_b$ is the barrier height. β, $\mu_b$, α, and $V_s$ are thermal conductivity, electron mobility in the barrier region, non-parabolicity coefficient, and the saturation velocity, respectively. Mobility in the well region depends on the doping concentration. Experimental measurements have been fit with the following equation in the range of practical dopings:
$\mu_w = 37666 - 1845.61 \times \log_{10}(n_e)$ in unit of cm$^2$/Vs, where $n_e$ is the bulk doping concentration in cm$^{-3}$.

Table I, J. M. O. Zide

1.  L. D. Hicks and M. S. Dresselhaus, Phys. Rev. B **47**, 12727 (1993).
2.  L. D. Hicks, T. C. Harman, and M. S. Dresselhaus, Appl. Phys. Lett. **63**, 3230 (1993).
3.  R. Venkatasubramanian, E. Siivola, T. Colpitts, and B. O'Quinn, Nature **413**, 597 (2001).
4.  T. C. Harman, P. J. Taylor, M. P. Walsh, and B. E. LaForge, Science **297**, 2229 (2002).
5.  A. Shakouri and J. E. Bowers, Appl. Phys. Lett. **71**, 1234 (1997).
6.  G. D. Mahan and L. M. Woods, Phys. Rev. Lett. **80**, 4016 (1998).
7.  C. B. Vining and G. D. Mahan, J. Appl. Phys. **86**, 6852 (1999).
8.  A. Shakouri, C. Labounty, P. Abraham, J. Piprek, and J. E. Bowers, Materials Research Society Proceedings **545**, 449 (1999).
9.  D. Vashaee and A. Shakouri, Phys. Rev. Lett. **92**, 106103/1 (2004).
10. D. Vashaee and A. Shakouri, J. Appl. Phys. **95**, 1233 (2004).
11. W. Kim, P. Reddy, A. Majumdar, J. Zide, A. Gossard, G. Zeng, J. Bowers, and A. Shakouri, in *ASME Integrated Nanosystems* (ASME, Pasadena, CA, U.S.A., 2004), p. NANO2004.
12. W. Kim, J. M. Zide, A. C. Gossard, D. O. Klenov, S. Stemmer, A. Shakouri, and A. Majumdar, Nature Materials **submitted** (2005).
13. D. O. Klenov, D. C. Driscoll, A. C. Gossard, and S. Stemmer, Appl. Phys. Lett. **86**, 111912 (2005).
14. A. Dorn, M. Peter, S. Kicin, T. Ihn, K. Ensslin, D. Driscoll, and A. C. Gossard, Appl. Phys. Lett. **82**, 2631 (2003).
15. C. Kadow, S. B. Fleischer, J. P. Ibbetson, J. E. Bowers, and A. C. Gossard, Appl. Phys. Lett. **75**, 3548 (1999).
16. P. Pohl, F. H. Renner, M. Eckardt, A. Schwanhausser, A. Friedrich, O. Yuksekdag, S. Malzer, G. H. Dohler, P. Kiesel, D. Driscoll, M. Hanson, and A. C. Gossard, Appl. Phys. Lett. **83**, 4035 (2003).
17. D. C. Driscoll, M. Hanson, C. Kadow, and A. C. Gossard, Appl. Phys. Lett. **78**, 1703 (2001).
18. D. G. Cahill, W. K. Ford, K. E. Goodson, G. D. Mahan, A. Majumdar, H. J. Maris, R. Merlin, and S. R. Phillpot, J. Appl. Phys. **93**, 793 (2003).
19. D. G. Cahill, K. Goodson, and A. Majumdar, J. Heat Transfer **124**, 223 (2002).
20. J. M. Zide, D. O. Klenov, S. Stemmer, A. C. Gossard, G. Zeng, J. E. Bowers, D. Vashaee, and A. Shakouri, Appl. Phys. Lett. **87**, 112102 (2005).
21. G. Zeng, J. E. Bowers, J. M. Zide, A. C. Gossard, Y. Zhang, A. Shakouri, W. Kim, S. Singer, and A. Majumdar, Appl. Phys. Lett. **submitted** (2005).